\documentclass{article}

\PassOptionsToPackage{numbers, compress}{natbib}

\usepackage[final]{neurips_2024}

\usepackage[utf8]{inputenc} 
\usepackage[T1]{fontenc}    
\usepackage{hyperref}       
\usepackage{url}            
\usepackage{booktabs}       
\usepackage{amsfonts}       
\usepackage{nicefrac}       
\usepackage{microtype}      
\usepackage{longtable} 
\usepackage{array}
\usepackage{geometry}
\usepackage{afterpage} 
\usepackage{subcaption} 
\usepackage{url}
\usepackage{graphicx}
\usepackage{adjustbox}
\usepackage{chngpage}

\usepackage{xcolor}
\usepackage{authblk}

\definecolor{red}{rgb}{1, 0, 0}

\title{CybORG++: An Enhanced Gym for the Development of Autonomous Cyber Agents}

\author[1,2]{Harry Emerson*}
\author[1]{Liz Bates*}
\author[1]{Chris Hicks}
\author[1]{Vasilios Mavroudis}

\affil[ ]{\texttt{harry.emerson@bristol.ac.uk, ebates@turing.ac.uk, c.hicks@turing.ac.uk, vmavroudis@turing.ac.uk}}
\affil[1]{The Alan Turing Institute, London, UK}
\affil[2]{University of Bristol, Bristol, UK}

\date{}

\begin{document}

\maketitle
\def\thefootnote{*}\footnotetext{These authors contributed equally to this work.}\def\thefootnote{\arabic{footnote}}

\begin{abstract}

CybORG++ is an advanced toolkit for reinforcement learning research focused on network defence. Building on the CAGE 2 CybORG environment, it introduces key improvements, including enhanced debugging capabilities, refined agent implementation support, and a streamlined environment that enables faster training and easier customization. Along with addressing several software bugs from its predecessor, CybORG++ introduces MiniCAGE, a lightweight version of CAGE 2, which improves performance dramatically—up to 1000× faster execution in parallel iterations—without sacrificing accuracy or core functionality. CybORG++ serves as a robust platform for developing and evaluating defensive agents, making it a valuable resource for advancing enterprise network defence research.\end{abstract} 

\section{Introduction}

As cyberattacks grow more frequent and sophisticated, the need for advanced, autonomous cyberdefence systems is critical. In this evolving landscape, reinforcement learning (RL) has emerged as a powerful tool for developing agents capable of protecting enterprise networks. Among the leading platforms in this domain, the CAGE 2 CybORG simulator~\cite{standen2021cyborg} stands out as a widely adopted environment for training and evaluating RL-based defensive agents. By simulating realistic network attack scenarios, CybORG has become a standard framework within the cybersecurity research community, providing a controlled yet dynamic environment for testing defence strategies.

However, as CybORG has gained traction, significant limitations have surfaced. Despite its robustness, the platform suffers from persistent software bugs and inefficiencies that undermine its reliability as a benchmark for RL agents. These issues, including erratic agent behavior caused by flawed code, often lead to inconsistent results, complicating reproducibility. Furthermore, CybORG’s architecture—designed with potential future emulation in mind—introduces unnecessary complexity, slowing down agent training and making it difficult for developers to debug agent actions. These challenges present barriers to progress in developing more capable autonomous defence agents.

In response to these issues, we introduce CybORG++, a comprehensive upgrade to the original CybORG platform. CybORG++ improves both the usability and reliability of the environment by addressing key bugs and streamlining agent interactions. It includes two major components: (1) a debugged version of the original CybORG CAGE 2, and (2) MiniCAGE, a lightweight re-implementation that preserves the core functionality while delivering significantly faster execution. MiniCAGE allows researchers to conduct large-scale experiments with up to 1000× speed improvements in parallel execution, drastically reducing the time required for training and evaluation. Additionally, an expanded developer guide offers in-depth documentation, providing greater transparency and ease of use for future research.

The CybORG++ toolkit represents a substantial improvement over its predecessor, offering researchers a more efficient and reliable platform for advancing autonomous cyber defence. By removing the bottlenecks that have hindered progress, CybORG++ facilitates faster experimentation, more accurate benchmarking, and greater confidence in RL-driven defence strategies, ultimately bringing the field closer to real-world applications of autonomous cybersecurity.



\section{CybORG - CAGE 2 Challenge}

The Cyber Operations Research Gym, or CybORG, is an open-source AI gym environment for developing blue and red team decision-making agents for cybersecurity tasks~\cite{standen2021cyborg, kiely2023autonomous}. The environment is designed to implement a variety of different scenarios, the most notable of which are the CAGE challenges~\cite{cage_challenge_1, kiely2023autonomous, cage_challenge_3_announcement, cage_challenge_4_announcement}. The CAGE challenges are a series of public challenges designed to foster the development of autonomous defensive agents. 

\begin{figure}[t!]
    \centering
    \includegraphics[width=0.9\textwidth]{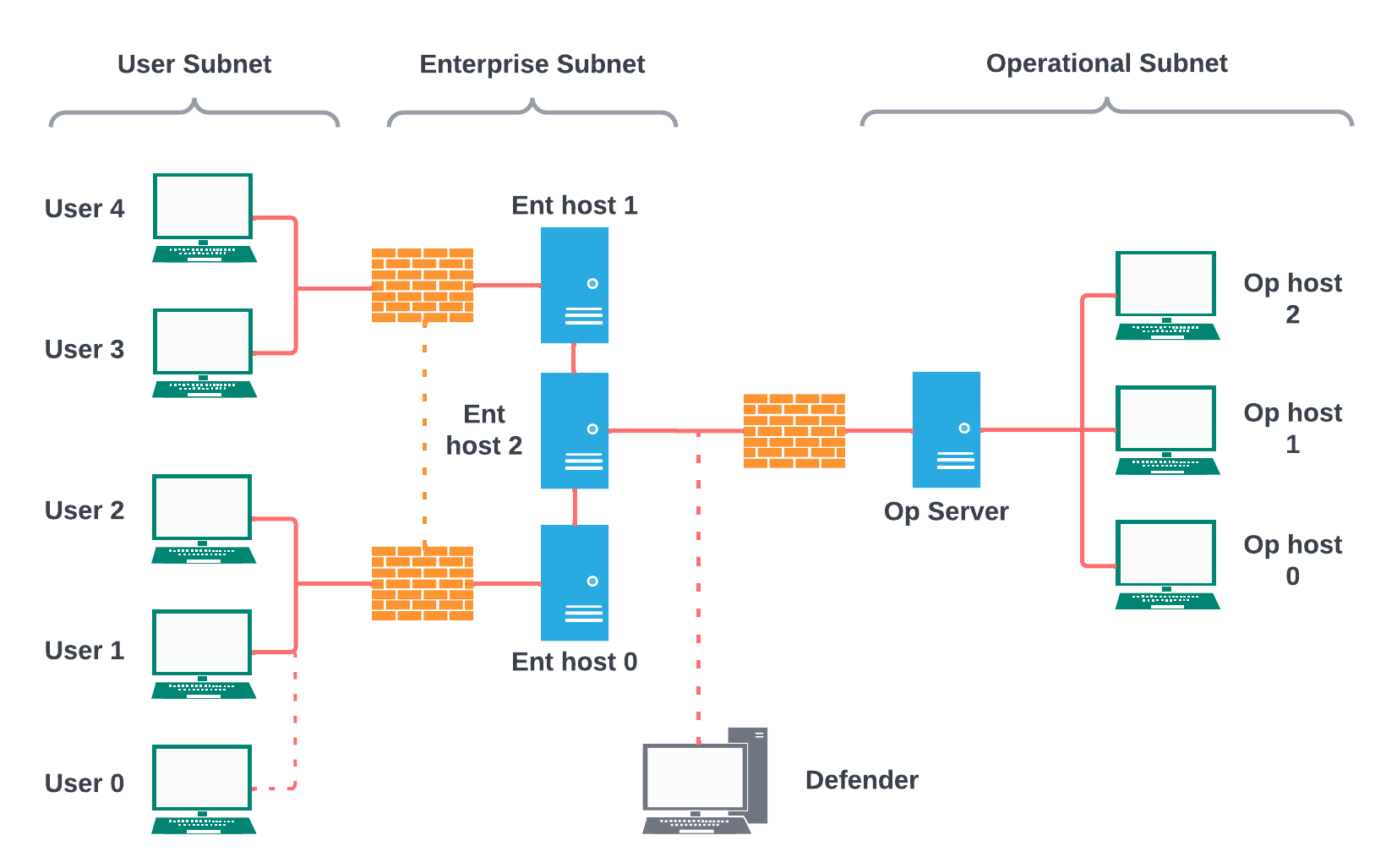} 
    \caption{CAGE 2 CybORG Network Diagram. The orange dotted line indicates a shared firewall between the User subnet and Enterprise subnet. The red dotted line indicates the defender is not a stationary host in the network, and that User0 is where red maintains a foothold on the system whilst not functioning as a proper user host. }
    \label{fig:Cage_network_diagram} 
\end{figure}

The first two CAGE challenges use a version of the CybORG environment which is compatible with training single defensive agents, whilst the third and fourth challenges focus on training multi-agent RL agents.  
The CAGE 2 Challenge is one of the most widely explored challenges~\cite{canaries2023,bates2023reward,foley2023inroadsautonomousnetworkdefence,vyas2023automated,foley2022autonomous,applebaum2022bridging} and focuses on developing a defensive blue agent able to defend a small enterprise network, seen in Figure \ref{fig:Cage_network_diagram}.

\subsection{Enhancements}

A variety of bugs were identified within the CAGE 2 CybORG environment, many of which affected core components of the environment and liked influenced defensive agent strategy. A detailed breakdown of the bugs can be found in the associated GitHub repository. The most notable are summarised as follows:
\begin{itemize}
    
    \item \textbf{Invalid Actions} - Several actions of the blue agent do not function as intended. For instance, the decoy for Vsftpd is incorrectly identified as the decoy for Apache, the fermitter decoy cannot be removed from a host, and certain hosts cannot be exploited by exploits that are purportedly compatible.

    \item \textbf{Inconsistent Reward Calculation} - The reward system for the blue agent is based on the number of hosts exploited; however, only specific types of exploits reliably trigger this reward.

    \item \textbf{Reduced Visibility in Observation} - In some cases, the blue agent's observations fail to detect red agent exploitation, resulting in a significantly lower detection rate for certain hosts.    
\end{itemize}

\subsection{Extended Developer Guide}

The CAGE 2 CybORG environment exhibits considerable complexity; consequently, the developer guide for the original implementation has been expanded to ensure that all the features of the environment remain fully transparent to reinforcement learning (RL) developers. This comprehensive guide includes an in-depth explanation of the state, action, and reward components, a detailed analysis of the pre-programmed behavior of the red agents, and an extensive reference detailing the interactions among various decoys, processes, and exploits across different hosts which can also be seen in the tables in appendix. There was not much other literature to cross reference our understanding of the system structure, though our perception of the decoy mappings and user processes and ports aligns with that of~\cite{hammar2024optimal}.


\section{MiniCAGE}

MiniCAGE is a lightweight version of CAGE 2 CybORG environment that enhances speed, transparency, and ease of use while staying true to the original challenge objectives. The package mimics the basic RL components of the original environment (e.g. state-action space, reward, etc.), but abstracts the bulky files and complex processes, resulting in a streamlined and more accessible framework that retains core functionalities. 

\subsection{Optimisations}

The main optimisations introduced by miniCAGE are the following:
\begin{itemize}
    
    \item \textbf{Parallel execution} - MiniCAGE supports parallel execution allowing agents to be trained and evaluated across thousands of different network configurations simultaneously on a single CPU. This modification accelerates mini-CAGE by almost 1000\(\times\) that of the original CybORG environment.
    
    \item \textbf{Red Agent Interface} - MiniCAGE can now be used to train both red and blue agents, enabling the exploration of more advanced offensive strategies compared to the pre-programmed agents in the original CybORG implementation.  
    
    \item \textbf{Simplified State-Action Space} - Environment inputs and outputs have been vectorised to more easily support RL training and evaluation. The default states from the original CybORG implementation have been enhanced with additional metrics to track malicious scans and decoy placements. The agent action space has been simplified. Decoy actions have been streamlined to a single action per host, automatically deploying the strongest decoy by default.

\end{itemize}

\subsection{Environment Benchmarking}

\begin{figure}[t!]
    \centering

    \begin{subfigure}[b]{0.45\textwidth}
        \centering
        \includegraphics[width=\textwidth]{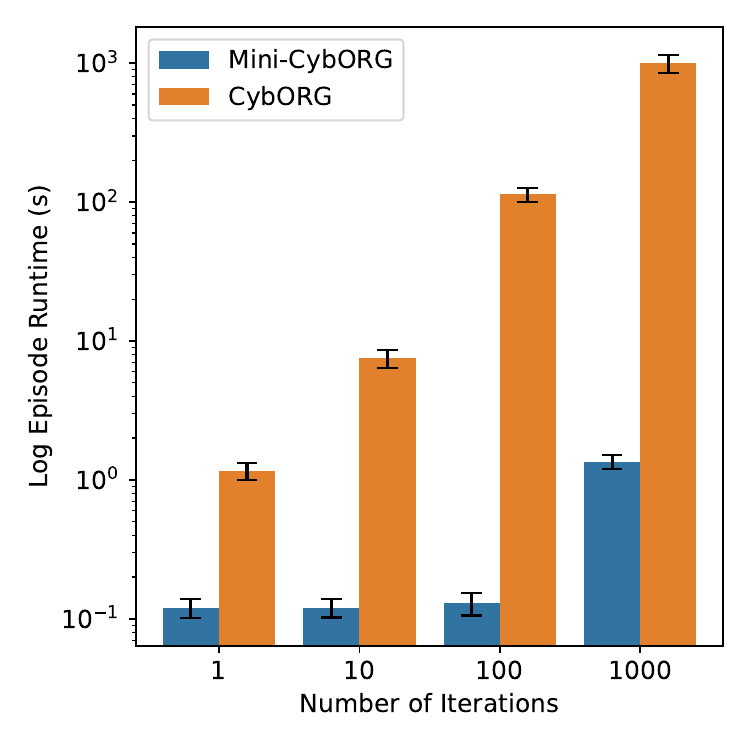}
        \caption{}
        \label{fig:speed_comparison}
    \end{subfigure}
    \begin{subfigure}[b]{0.45\textwidth}
        \centering
        \includegraphics[width=\textwidth]{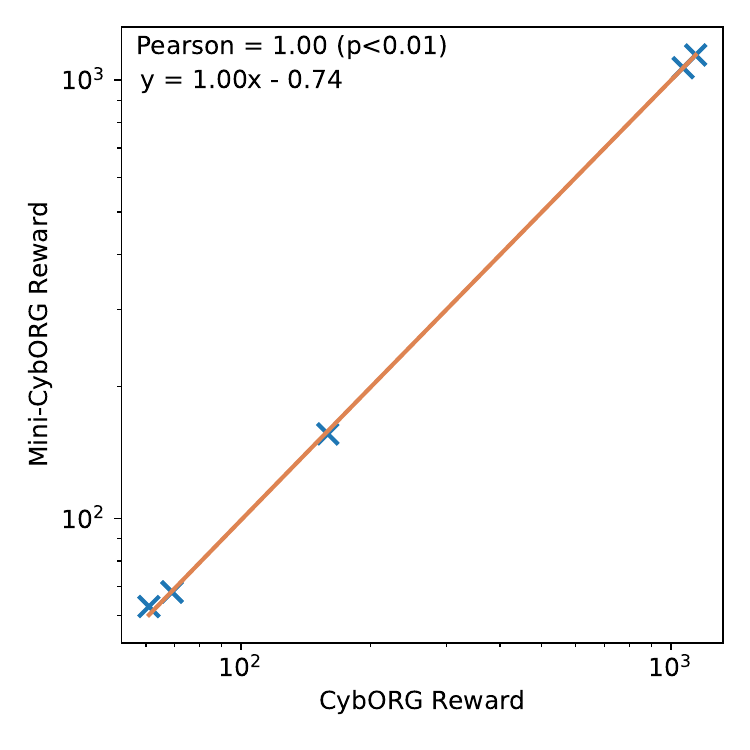}
        \caption{}
        \label{fig:reward_comparison}
    \end{subfigure}

    \caption{Comparison of the developed MiniCAGE environment to the original CAGE 2 CybORG implementation. a) Execution speed improvement of MiniCAGE compared to CAGE 2 CybORG environment. MiniCAGE is highlighted to run is approximately 950\(\times\) faster than CAGE 2 CybORG when running over 100 parallel iterations on a single CPU. Error bars show the standard error. b) Performance of six attacker-defender pairs in both environments to confirm the equivalence of the CybORG CAGE 2 and MiniCAGE environments. The strong correlation in agent behavior observed between both environments indicates consistent environmental dynamics.}    
    \label{fig:cage_comparison}
\end{figure}

To explore the speed improvement resulting from MiniCAGE, both environments were run over a specified number of iterations with random starting initialisations and input actions. The time elapsed between initialisation and completion was then used as a metric of environmental execution speed. The results were aggregated over 100 separate instances. MiniCAGE offers significant speed improvements over the original CybORG CAGE 2 environment, as evidenced by Figure~\ref{fig:speed_comparison}. Over the four iteration intervals, MiniCAGE offers speed improvements of approximately \(15\times\), \(65\times\), \(950\times\) and \(800\times\), respectively.

To verify that MiniCAGE accurately emulates the original CAGE 2 environment, six defender-attacker pairs were tested (defenders: react-restore, react-decoy, sleep; attackers: b-line, meander). Each agent employs a distinct strategy, ensuring that, together, the agents explore the full state-action space. Each agent pair was repeated for 500 episodes for 100 timesteps each. The reward for the agent pairs was summed for each episode and used as a metric for agent behaviour. Figure \ref{fig:reward_comparison} shows the equivalence of the two environments and demonstrates a strong correlation in reward (Pearson correlation of 1.00 (p \(< 0.01\)) and all reward measurements fall well within the standard error of the comparison environment.


\section{Potential Future Directions}

\begin{itemize}
    
    \item \textbf{Community Contribution} - CybORG++ is the product of an interdisciplinary effort aimed at reigniting interest in the application of reinforcement learning to cybersecurity and network defence. We actively encourage community contributions to the repository, particularly those that focus on identifying bugs in the CAGE 2 CybORG environment or enhancing the functionality of MiniCAGE.

    \item \textbf{Extension of MiniCAGE} - MiniCAGE's streamlined design makes it highly adaptable for modifications and improvements. Future enhancements could include implementing custom network scenarios within the simulator, particularly those replicating more realistic network configurations. Additionally, re-writing the environment in JAX could likely lead to further performance improvements.

    \item  \textbf{Comprehensive Benchmarking} - Benchmarking within CybORG++ could be enhanced by establishing a leaderboard that tracks performance across various cyber defence scenarios. This could also involve storing the top-performing models, along with their code and weights, within the repository.

\end{itemize}

\section{Codebase}
For access to the MiniCAGE environment and a separate debugged version of the existing CAGE 2 CybORG environment, please see the following repository:\\
\hspace{1ex}
\url{https://github.com/alan-turing-institute/CybORG_plus_plus}

\section{Acknowledgements}
Research funded by the Defence Science and Technology Laboratory (DSTL) which is an executive agency of the UK Ministry of Defence providing world class expertise and delivering cutting-edge
science and technology for the benefit of the nation and allies. The  research supports the Autonomous Resilient Cyber Defence (ARCD) project within the Dstl Cyber Defence Enhancement programme.

\bibliographystyle{plain} 
\bibliography{references} 

\section{Appendix}

{
\thispagestyle{plain}

{
\setlength\LTleft{-0.5in}

\begin{longtable}{@{}l l >{\raggedright\arraybackslash}p{2cm} p{2.5cm} p{3.0cm} p{4.5cm}@{}}
\caption{Host Details with Ports, Decoys, Decoy Order, and Initial Exploit Order} \\ 
\toprule
\textbf{Host} & \textbf{OS} & \textbf{Local Ports} & \textbf{Decoys} & \textbf{Decoy Order} & \textbf{Initial Exploit Order} \\
\midrule
\endfirsthead
\caption[]{(Continued) Host Details with Ports, Decoys, Decoy Order, and Initial Exploit Order} \\ 
\toprule
\textbf{Host} & \textbf{OS} & \textbf{Local Ports} & \textbf{Decoys} & \textbf{Decoy Order} & \textbf{Initial Exploit Order} \\
\midrule
\endhead
\textbf{User0} & Windows & 21, 22 & Apache,\newline Smss,\newline Svchost,\newline Tomcat & Svchost - 1\newline Smss - 2\newline Apache - 3\newline Tomcat - 4 & FTPDirTraversal (port 21) : 7,\newline SSHBruteForce (port 22) : 0.1 \\
\textbf{User1} & Windows & 21, 22 & Apache,\newline Smss,\newline Svchost,\newline Tomcat & Svchost - 1\newline Smss - 2\newline Apache - 3\newline Tomcat - 4 & FTPDirTraversal (port 21) : 7,\newline SSHBruteForce (port 22) : 0.1 \\
\textbf{User2} & Windows & 445, 139, 135, 3389 & Apache,\newline SSHd,\newline Tomcat,\newline Femitter & SSHd - 0.1\newline Apache - 3\newline Tomcat - 4\newline Femitter - 7 & EternalBlue (port 139) : 2,\newline BlueKeep (port 3389) : 1 \\
\textbf{User3} & Linux & 25, 80, 443, 3390 & SSHd,\newline Vsftpd & SSHd - 0.1\newline Vsftpd - 7 & HarakaRCE (port 25) : 6,\newline SQLInjection (port 3390 and 80/443) : 5,\newline HTTPsRFI (port 443) : 4,\newline HTTPRFI (port 80) : 3,\newline BlueKeep (port 3389) : 1 \\
\textbf{User4} & Linux & 22, 80, 3390, 443, 25 & Vsftpd & Vsftpd - 7 & HarakaRCE (port 25) : 6,\newline SQLInjection (port 3390 and 80/443) : 5,\newline HTTPsRFI (port 443) : 4,\newline HTTPRFI (port 80) : 3,\newline BlueKeep (port 3389) : 1 \\
\textbf{Ent0} & Linux & 22 & Apache,\newline Tomcat,\newline Vsftpd,\newline HarakaSMPT & Apache - 3\newline Tomcat - 4\newline Vsftpd - 7\newline HarakaSMPT - 6 & SSHBruteForce (port 22) : 0.1 \\
\textbf{Ent1} & Windows & 22, 135, 3389, 445, 139, 80, 443 & Femitter & Femitter - 7 & HTTPsRFI (port 443) : 4,\newline HTTPRFI (port 80) : 3,\newline EternalBlue (port 139) : 2,\newline BlueKeep (port 3389) : 1,\newline SSHBruteForce (port 22) : 0.1 \\
\textbf{Ent2} & Windows & 22, 135, 3389, 445, 139, 80, 443 & Femitter & Femitter - 7 & SSHBruteForce (port 22) : 0.1 \\

\pagebreak

\textbf{Op\_host0} & Linux & 22 & Apache,\newline HarakaSMPT,\newline Tomcat,\newline Vsftpd & Vsftpd - 7\newline HarakaSMPT - 6\newline Tomcat - 4\newline Apache - 3 & SSHBruteForce (port 22) : 0.1 \\
\textbf{Op\_host1} & Linux & 22 & Apache,\newline HarakaSMPT,\newline Tomcat,\newline Vsftpd & Vsftpd - 7\newline HarakaSMPT - 6\newline Tomcat - 4\newline Apache - 3 & SSHBruteForce (port 22) : 0.1 \\
\textbf{Op\_host2} & Linux & 22 & Apache,\newline HarakaSMPT,\newline Tomcat,\newline Vsftpd & Vsftpd - 7\newline HarakaSMPT - 6\newline Tomcat - 4\newline Apache - 3 & SSHBruteForce (port 22) : 0.1 \\
\textbf{Op\_Server} & Linux & 22 & Apache,\newline HarakaSMPT,\newline Tomcat,\newline Vsftpd & Vsftpd - 7\newline HarakaSMPT - 6\newline Tomcat - 4\newline Apache - 3 & SSHBruteForce (port 22) : 0.1 \\
\bottomrule
\end{longtable}
}

\begin{longtable}{@{}l p{4cm} p{4cm}@{}}
\caption{Exploits, Decoys, and Processes} \\ 
\toprule
\textbf{Exploits} & \textbf{Decoys} & \textbf{Processes} \\
\midrule
\endfirsthead
\caption[]{(Continued) Exploits, Decoys, and Processes} \\ 
\toprule
\textbf{Exploits} & \textbf{Decoys} & \textbf{Processes} \\
\midrule
\endhead
EternalBlue & decoySmss & smss.exe \\
BlueKeep & decoySvchost & svchost.exe \\
HTTPRFI & decoyApache & apache2 \\
HTTPSRFI & decoyTomcat & tomcat8.exe \\
SSH BruteForce & decoySSHD & sshd.exe/sshd \\
SQL Injection & - & mysql \\
Haraka RCE & decoyHarakaSMTP & smtp \\
FTP Directory Traversal & decoyFemitter\newline decoyVsftpd & femitter.exe \\
\bottomrule
\end{longtable}

\begin{longtable}{@{}l p{3cm} p{4cm} p{4cm}@{}}
\caption{Hostname, Ports, Users, and Processes} \\ 
\toprule
\textbf{Hostname} & \textbf{Ports} & \textbf{Users} & \textbf{Processes} \\
\midrule
\endfirsthead
\caption[]{(Continued) Hostname, Ports, Users, and Processes} \\ 
\toprule
\textbf{Hostname} & \textbf{Ports} & \textbf{Users} & \textbf{Processes} \\
\midrule
\endhead
User1 & 22,\newline 21 & SSHD\_SERVER, \newline SYSTEM & SSHD.EXE,\newline FEMITTER.EXE \\
User2 & 445, 139,\newline 135,\newline 3389 & SYSTEM,\newline SYSTEM,\newline NETWORK & SMSS.EXE,\newline SVCHOST.EXE,\newline SVCHOST.EXE \\
User3 & 3389,\newline 80, 443,\newline 25 & ROOT,\newline WWW-DATA,\newline ROOT & MYSQL,\newline APACHE2,\newline SMTP \\
User4 & 22,\newline 3390,\newline 80, 443,\newline 25 & ROOT,\newline ROOT,\newline WWW-DATA,\newline ROOT & SSHD,\newline MYSQL, APACHE2,\newline SMTP \\
Ent0 & 22 & ROOT & SSHD.EXE \\
Ent1 & 22,\newline 135,\newline 3389,\newline 445, 139,\newline 80, 443 & SSHD\_SERVER,\newline SYSTEM,\newline SYSTEM,\newline SYSTEM,\newline NETWORK & SSHD.EXE,\newline SVCHOST.EXE,\newline SVCHOST.EXE,\newline SMSS.EXE,\newline TOMCAT8.EXE \\
Ent2 & 22,\newline 135,\newline 3389,\newline 445, 139,\newline 80, 443 & SSHD\_SERVER,\newline SYSTEM,\newline SYSTEM,\newline SYSTEM,\newline NETWORK & SSHD.EXE,\newline SVCHOST.EXE,\newline SVCHOST.EXE,\newline SMSS.EXE,\newline TOMCAT8.EXE \\
Op\_Server & 22 & ROOT & SSHD \\
Op\_host0 & 22 & ROOT & SSHD \\
Op\_host1 & 22 & ROOT & SSHD \\
Op\_host2 & 22 & ROOT & SSHD \\
Defender & 22,\newline 53, 78 & ROOT, \newline SYSTEMD+ & SSHD,\newline SYSTEMD \\
\bottomrule
\end{longtable}
}
\restoregeometry

\end{document}